
\input jnl
\input eqnorder
\input reforder
\preprintno{OUTP--93--41S}
\def\Jb{\bar J}
\def\partialb{\bar\partial}
\def\zb{{\bar z}}
\def\no#1{:\!#1\!:}
\def\sec#1{Sec.\thinspace#1}
\title
Continuously Varying Exponents for Oriented Self-Avoiding Walks
\author
John Cardy
\affil
All Souls College\\{\rm and}\\Department of Physics\\
Theoretical Physics\\1 Keble Road\\Oxford OX1 3NP, UK
\abstract
A two-dimensional conformal field theory with
a conserved $U(1)$ current $\vec J$, when perturbed by the 
operator ${\vec J}^{\,2}$, exhibits a line of fixed points along which 
the scaling dimensions of the operators with non-zero $U(1)$ charge vary
continuously. This result is applied to the problem of oriented
polymers (self-avoiding walks) in which the short-range repulsive
interactions between two segments depend on their relative orientation.
While the exponent $\nu$ describing the fractal dimension of such walks
remains fixed, the exponent $\gamma$, which gives the total number
$\sim N^{\gamma-1}\mu^N$ of such walks, is predicted to vary continuously
with the energy difference.

\endtitlepage
\oneandahalfspace

\head{1. Introduction.}

Critical exponents which change continuously as a parameter of the
underlying hamiltonian is varied are, by now, a well-understood
phenomenon in the theory of critical behaviour. In the framework of the
renormalisation group, they correspond to the existence of continuous
manifolds of fixed points, rather than isolated points. In the earliest
examples found, {\it e.g.} the low-temperature phase of the two-dimensional
XY model, and the critical line of the Ashkin-Teller 
model, these manifolds are one-dimensional,
but, with the development of conformal
field theory (CFT), it was realised that it is possible for the moduli
space of such fixed points to be multi-dimensional.
The theorem of Friedan, Qiu and Shenker\refto{FQS} shows that, in a unitary 
CFT, continuously varying exponents cannot arise when the central charge
$c<1$. Indeed, the examples alluded to above are now known\refto{NIEN} to
correspond to a $c=1$ theory, that of
free scalar field compactified on a circle, with action
$$
S_G={g\over4\pi}\int(\partial\phi)^2d^2z
\eqno(G)
$$ 
Unitary theories with higher dimensional moduli spaces of fixed points
have larger values of
$c$. While these are of great interest for string theory, their
applicability to critical phenomena appears limited.

In this paper we point out that continuously varying exponents also may
arise in theories with $c<1$, as long as they are non-unitary. In
particular, they occur in the $O(2n)$ model, with $-1\leq n\leq1$, which
has a statistical interpretation as a gas of loops, and whose $n\to0$
limit corresponds to the problem of self-avoiding walks (SAW). 
The main point is that a sufficient condition for a given CFT to be
part of a larger moduli space of such theories is that it possess
a conserved current, with complex components $(J,\Jb)$,
which generates a $U(1)$ symmetry of the theory.
For, in that case, as will be shown, when a term $\lambda\int\no{J\Jb}
d^2z$
is added to the action, the perturbation is `truly' marginal in that
its renormalization group flow vanishes to all orders, and it therefore
generates a line of fixed points. In addition,
the scaling dimensions $x_q$ of those fields
of the unperturbed theory which have non-zero $U(1)$ charge $q$ then become
continuously dependent on $\lambda$, according to the formula
$$
x_q(\lambda)=x_q(0)+{2\pi\lambda q^2\over 1-2\pi k\lambda}
\eqno(XL)
$$
where $k$ is the `chiral anomaly', given by the coefficient of the
correlation function\break $\langle J(z)J(0)\rangle=k/z^2$.

This result is in fact a special case of a more general result of 
Chaudhuri and Schwartz\refto{CS}, who showed that, for a general 
current $J$, a sufficient condition that $J\Jb$ generate a truly marginal
perturbation is that the OPE  $J(z)J(0)$ should contain no simple pole term
proportional to $z^{-1}$. For the special case of a $U(1)$ current, this
condition is satisfied by symmetry. 

For the Gaussian model \(G), the current components $J$ and $\Jb$ are
simply proportional to $\partial\phi$ and $\partialb\phi$ respectively, so 
that the marginal perturbation simply corresponds to changing the radius 
parameter $g$. One may, however, also derive the marginality explicitly by
considering perturbation theory in $\lambda$. Each order is given in
terms of a sum of integrals over correlation functions of the
unperturbed theory. These may be regulated at short distances by a
suitable explicit cut-off, and at large distances, if necessary,
by considering the theory on the cylinder. While the explicit integrals
become increasingly difficult to evaluate, one knows that they must
resum to give the expected result. Now, in the case of a more general
theory, perturbed by $J\Jb$, the $U(1)$ Ward identities completely fix the
form of the correlation functions with an arbitrary number of insertions
of this operator. Apart from a trivial relabelling of the coupling
constant and the charges, these become identical with those of the
$c=1$ theory. Therefore we need do no more work in evaluating them:
the existence of a line of fixed points and the result \(XL) for the
variation of the scaling dimensions follow immediately.

When applied to CFTs based on models with a symmetry group which
contains $U(1)$ as a subgroup, for example $SU(N)$ WZWN models, this
result is not new: is gives a part of the well-known moduli space of
such theories.
However, there exist also CFTs with $c<1$ which possess conserved
$U(1)$ currents. The $O(2n)$ model referred to above may be more
profitably viewed as a {\it complex} $O(n)$ model. In that guise, it
is equivalent to a gas of {\it oriented} loops, in which both
orientations are summed over with equal weights. Examination of the
partition function of this model on, for example, honeycomb lattice,
shows that the loops are self-avoiding, with a fugacity $n$ for each
loop, and that their interactions, in this unperturbed model, are
independent of their orientation. Now consider putting a unit current
around each loop, in the direction of its orientation\refto{MILL}. This 
corresponds in the complex $O(n)$ model to a conserved current
whose integral generates
the $U(1)$ phase symmetry of the complex-valued fundamental fields.
In this language, the marginal perturbation $J\Jb$ is the continuum
limit of an short-range interaction $\lambda_0$ between neighboring loops which
now depends on their relative orientation. In two dimensions, two
segments of an oriented self-avoiding walk may approach each other 
in two ways: their respective orientations may either be parallel or
anti-parallel (see \fig{1}). Adding such a $J\Jb$ term to the action
corresponds to weighting these two possibilities by different amounts.
The general result described above then implies that the scaling
dimensions of operators with non-zero $U(1)$ charge will depend
continuously on this energy difference $\lambda_0$. 
These operators will act as
sources and sinks for the current, and therefore correspond to the
end-points of such oriented self-avoiding walks (SAWs). 
Thus we predict that that the
exponent $\gamma$, which relates to the total number $c_N$
of open SAW of length $N$ via the formula $c_N\sim N^{\gamma-1}\mu^N$,
will depend continuously on the energy difference. 
Similarly, one may consider `star' oriented polymers, formed by the
joining of $q$ oriented SAW at their sources.
The corresponding exponent $\gamma_q$ also depends continuously on
$\lambda_0$, according to the law
$$
\gamma_q(\lambda)=\gamma_q(0)-2\pi q(q+1)\lambda
\eqno(GQ)
$$
where $\lambda$ is a non-universal monotonic function of $\lambda_0$.
Thus, although the various exponents are non-universal, through
elimination of $\lambda$ there do exist
universal relations between them, as with other examples of
one-dimensional manifold of fixed points.

On the other hand,
the exponent $\nu$, which determines the fractal dimension $\nu^{-1}$
of SAWs, is
related to the scaling dimension of the energy operator, which has zero
charge and therefore should remain constant. This is consistent with the
observation that for self-avoiding loops, whose statistics are
determined by correlations of energy operators,
only anti-parallel contacts are
possible, and they are therefore insensitive to the perturbation.
(Merely changing the interaction energy for anti-parallel contacts is
believed to correspond to an irrelevant perturbation, as long as it
remains repulsive.)

The outline of this paper is follows. In \sec{2} we first analyse the
perturbation theory for a marginal perturbation in the $c=1$ theory. We
then generalise to that of a general model with a $J\Jb$ perturbation,
and show how the perturbative expansion is simply related to that of the
case $c=1$. Making this correspondence, we deduce the relation \(XL).
In the second part of this Section, we show how this shift in the
scaling dimensions may be explained in terms of a deformation of the
stress tensor, and show that nevertheless the central charge $c$ is not
affected, as expected from Zamolodchikov's $c$-theorem\refto{ZAM}.
While this shows how things work from the point of view of CFT, this
subsection is not necessary for the subsequent development and may be
omitted.
In \sec{3} we discuss how all this applies to the $O(n)$ model, and
in particular the dilute limit $n\to0$. Finally, in \sec{3} we make some
concluding remarks.

\head{2. Analysis of $J\Jb$ perturbation.}

We first discuss in detail the example of the $c=1$ Gaussian model
\(G). All correlation functions of this theory are related by Wick's
theorem to that
of the fundamental field $\langle\phi(z,\zb)\phi(0)\rangle=
-(1/g)(\ln z+\ln\zb)$. 
In particular, the two-point function of the operator $V_q\equiv
\no{e^{iq\phi}}$ has the form $\langle V_q(r)V_{-q}(0)\rangle
\sim r^{-q^2/g}$
so that this operator has scaling dimension $x_q(g)=q^2/2g$.
The theory possesses a $U(1)$ invariance corresponding to constant
shifts in $\phi$, and the current generating this symmetry is
$J_\mu=2ig\partial_\mu\phi$. Its normalization is fixed by the 
operator product expansion (OPE)
of $J$ with the operator $V_q$, which has $U(1)$ charge $q$:
$$
J(z)\cdot V_q(0)={q\over z}V_q(0)+\cdots
\eqno(JOPE)
$$
In this normalization, the $U(1)$ charge is $(1/2\pi)\int J_0dx^1$,
with the factor of $(1/2\pi)$ conventional to conformal field theory.

Now set $g=g_0$, corresponding to the unperturbed theory, and
consider the effect on the correlation function 
$G_q(r)=\langle V_q(r)V_{-q}(0)\rangle$ of adding a term 
$\delta S=\lambda\int \no{J\Jb}d^2z$ to $S_G$.\footnote*{Normal ordering
is defined in the standard way by point-splitting and subtracting the
singular terms in the OPE. This depends on the coupling constant $g$,
and, by convention, all normal ordering is defined in the unperturbed
theory.} 
The result may be expressed as a perturbative expansion in $\lambda$
of the form
$$
G_q(r)=\sum_nG^{(n)}_q(r)\lambda^n
\eqno(PT)
$$
where $G_q^{(n)}$ is given by an integral over the connected correlation
function with $n$ insertions of $J\Jb$. For example
$$
G^{(1)}_q(z_1-z_2)=4g_0^2\int\langle
V_q(z_1,\zb_1)V_{-q}(z_2,\zb_2)\no{\partial\phi(z)\partialb\phi(\zb)}
\rangle d^2z
\eqno()
$$
Using Wick's theorem, there are four terms, corresponding to the cases
where $\phi(z)$ and $\phi(\zb)$ are contracted onto $V_q$ and $V_{-q}$
respectively. After some simple algebra, these may be combined to give
$$
-q^2\int{|z_1-z_2|^2\over|z-z_1|^2|z-z_2|^2}d^2z
\eqno()
$$
As it stands, this integral is in need of ultraviolet regularization,
which may be implemented with a simple cut-off $|z-z_i|>a$. The
resulting singular dependence is then of the form
$$
-4\pi q^2\ln(|z_1-z_2|/a)
\eqno()
$$
The dependence on $a$ may be absorbed into an extra multiplicative
renormalization of the operators $V_{\pm q}$. The dependence on 
$|z_1-z_2|$ is consistent with a shift in the scaling dimension
$$
x_q={q^2\over2g_0}+2\pi\lambda q^2+O(\lambda^2)
\eqno(DG)
$$
Of course, we could have obtained this another way, by simply observing
that the perturbation corresponds to a shift in the coupling constant
$$
g_0\to g=g_0-4\pi\lambda g_0^2
\eqno(SHIFT)
$$
which is consistent with \(DG).

The next order contribution is proportional to
$$
\int\langle
V_q(z_1,\zb_1)V_{-q}(z_2,\zb_2)\no{\partial\phi(z)\partialb\phi(\zb)}
\no{\partial\phi(z')\partialb\phi(\zb')} \rangle_c d^2zd^2z'
\eqno()
$$
The different kinds of Wick contraction are now illustrated in \fig{2}.
Those of type (a) give a further multiplicative renormalization of
$V_{\pm q}$. Diagram (b) is clearly proportional to the square of the
first order contribution, and corresponds to the next term in the
exponentiation of \(DG). Contributions of type (c) are new, and involve
the correlation function $\langle J(z)J(z')\rangle=k_G/z^2$, where
$k_G=2g_0$. Explicit evaluation of these integrals, in the limit
$|z_1-z_2|\gg a$, is straightforward but lengthy. They give rise to
a contribution proportional to $\lambda^2q^2r^{-q^2/g_0}\ln(r/a)$,
where $r=|z_1-z_2|$. However, it is not necessary to perform the
integrals explicitly, since we know from \(SHIFT) that such terms must
come from the next order in the expansion of
$$
x_q(g)={q^2\over2(g_0-4\pi\lambda g_0^2)}
={q^2\over2g_0}+2\pi\lambda q^2+8\pi^2\lambda^2g_0q^2+\cdots
\eqno(XG)
$$

Now consider the case of a general theory with a $U(1)$ current,
perturbed by $\lambda J\Jb$. Let us consider again the 
perturbative expansion of the two-point function $\langle V_q(z_1)
V_{-q}(z_2)\rangle$, where now $V_q$ represents any primary
operator of charge
$q$ with respect to the chosen $U(1)$ symmetry. Note that, in a given
CFT, there may be more than one such operator: in that case, we choose
a basis such that 
$\langle V^i_q(z_1)V^j_{-q}(z_2)\rangle\propto\delta_{ij}$.
It is then straightforward to show that the perturbation does not mix
these operators. Note also, that in a general theory, the unperturbed
scaling dimension $x_q(\lambda=0)$ is not necessarily proportional to
$q^2$. The perturbation expansion is once again given in terms of
integrals over connected correlation functions with arbitrary numbers
of insertions of $J\Jb$. The main point is now that these are
completely determined by the $U(1)$ Ward identities and the OPE
$$
J(z)J(z')={k\over(z-z')^2}+O\big((z-z')^0\big)
\eqno(JJOPE)
$$
where the $U(1)$ anomaly number $k$ is fixed by the theory. This result
is most easily proved by induction. Consider the correlation function
with $n$ insertions of $J$
$$
\langle J(z)J(z_2)\ldots J(z_n)\prod_jV_{q_j}(\zeta_j)\rangle
\eqno()
$$
Regarded as function of $z$, this is analytic apart from simple poles
at $z=\zeta_j$, whose residue is given by the OPE 
$J(z)V_{q_j}(\zeta_j)=(q_j/z-\zeta_j)V_{q_j}(\zeta_j)+\cdots$,
and double poles at $z=z_i$, whose coefficient is given by \(JJOPE).
In addition, this function must behave as $z^{-2}$ as $z\to\infty$,
as may be seen from its covariance under the conformal transformation
$z\to z^{-1}$. Thus it is completely determined by the coefficients
of its singular behaviour, and these, in turn, are determined by 
similar correlation functions with lower values of $n$.

A similar result applies to insertions of $\Jb$.
For simultaneous insertions of $J$ and $\Jb$, we also need the contact
term
$$
J(z)\Jb(\zb')=(\pi k/2)\delta^{(2)}(z-z')
\eqno()
$$
which follows from \(JJOPE) and the conservation law $\partialb J+
\partial\Jb=0$.
We see then that all correlation functions with arbitrary
insertions of $J$ and $\Jb$ are identical with those of the Gaussian model,
with the sole exception that $k_G=2g_0$ is replaced by $k$.
The same is therefore true of $\no{J\Jb}$ insertions.
Thus these integrals may be evaluated with no work. 
The scaling dimensions of operators with charge $q$ are
modified. The shift may be determined by writing 
\(XG) as
$$
x_q(\lambda)={q^2\over 2(g_0-4\pi\lambda g_0^2)}
=x_q(0)+{2\pi\lambda q^2\over 1-4\pi\lambda g_0}
\eqno()
$$
and simply making the replacement $g_0\to k/2$ in the denominator
of the second term. This is because, as we argued above, the higher
order terms in the expansion all come from diagrams where some of
the $J$s are contracted onto other $J$s, and this involves the actual
$U(1)$ anomaly $k$ of the theory, rather than $2g_0$. This then gives the
main result \(XL) quoted in the Introduction.

\subhead{Perturbed stress tensor.}
This result may also be seen in terms of a modification of the stress
tensor $T(z)$ of the original theory.
As above, we first analyse this in the $c=1$ theory. The stress tensor
of the Gaussian model is $T(z)=-g\no{(\partial\phi)^2}$. The
normalization of this may be checked by considering the OPE with
$V_q$, for example, which gives 
$$
T(z)\cdot V_q(0)={q^2/4g\over z^2}V_q(0)+\cdots
\eqno(TV)
$$
consistent with the scaling dimension $x_q=q^2/2g$. As before let us
write this as
$$
T=-(g_0-4\pi\lambda g_0^2)\no{(\partial\phi)^2}
\equiv T_0-\pi\lambda\no{J^2}
\eqno(TP)
$$
To lowest order in $\lambda$, we have the OPEs
$$
\eqalign{
T_0\cdot V_q&={q^2/4g_0\over z^2}V_q+\cdots\cr
\no{J^2}\cdot V_q&={q^2\over z^2}V_q+\cdots\cr}
\eqno(TOV)
$$
but there are also contributions of higher order in $\lambda$ to the
$O(z^{-2})$ terms which come from insertions of $J\Jb$. Once again,
the integrals involved rapidly become tedious to compute, but we know
that they must sum to the result in \(TV).
Therefore, to all orders in $\lambda$, \(TOV) is replaced by
$$
\eqalign{
T_0\cdot V_q&={q^2/4g_0\over (1-4\pi\lambda g_0)^2z^2}V_q+\cdots\cr
\no{J^2}\cdot V_q&={q^2\over (1-4\pi\lambda g_0)^2z^2}V_q+\cdots\cr}
\eqno()
$$
so that, combining these, we find
$$
T\cdot V_q=(T_0-\pi\lambda\no{J^2})\cdot V_q=
{q^2\over(g_0-4\pi\lambda g_0^2)z^2}V_q+\cdots
\eqno(TVQ)
$$
consistent with the known shift in the scaling dimension.

For a more general theory perturbed by $\lambda J\Jb$ this result
may now be taken over simply, with $(q^2/4g_0)$ replaced by $x_q(0)/2$,
and the denominator $1-4\pi\lambda g_0$ replaced by $1-2\pi\lambda k$.
For example, consider the $O(\lambda)$ correction to $T_0\cdot V_q$,
which involves the integral
$$
\int T_0(z)J(z_1)\Jb(z_1)V_q(0)d^2z_1
\eqno()
$$
The important contributions come from the contact term in the
contraction of $T$ with $\Jb$:
$$
T_0(z)\Jb(z_1)=(\pi/2)\delta^{(2)}(z-z_1)J(z_1)
\eqno()
$$
which follows from the OPE $T\cdot J$ and current conservation. Thus this term
is proportional to $q^2$ with a universal coefficient. However, the
higher order terms in $\lambda$ all involve contractions of $J$ with
$J$. and therefore depend also on $k$. The form of the full answer may
be seen by writing the coefficient of $(1/z^2)V_q$ on the right hand
side of \(TVQ) as
$$
{q^2\over 4g_0}+
{2\pi\lambda-4\pi^2\lambda^2g_0\over(1-4\pi\lambda g_0)^2}q^2
\eqno()
$$
then replacing the first term by $x_q(0)$, and $g_0$ in the last term
by $k/2$. 
A similar result holds for $\no{J^2}\cdot V_q$. By construction, the
complicated terms proportional to $q^2$ will then simplify in the sum
$(T_0-\pi\lambda\no{J^2})\cdot V_q$, with a result consistent with
\(XL).

A similar calculation may be done for the central charge of the
perturbed theory, defined as the coefficient $c$ of the two-point
function $\langle T(z)T(0)\rangle=c/(2z^4)$. 
According to Zamolodchikov's $c$-theorem\refto{ZAM}, this should 
not change along a manifold of fixed points.
Writing $T=T_0-\pi\lambda\no{J^2}$ as before, in the Gaussian model
we have
$$
\langle T_0(z)T_0(0)\rangle={g_0^2\over g^2}{1\over 2z^4}=
{1\over(1-4\pi\lambda g_0)^2}{1\over2z^4}
\eqno(TT)
$$
In the general case, the $O(\lambda^0)$ term is of course proportional
to the central charge $c_0$ of the unperturbed theory. However, the
$O(\lambda)$ term arises from an integration over the connected part of
$\langle TT\no{J^2}\rangle$. Since only the connected part is included,
this does not depend on $c_0$, but only the (universal) scaling
dimension of $\no{J^2}$. Similarly, all the other higher order terms
may involve $\langle JJ\rangle$ contractions, but not $\langle
T_0T_0\rangle$, and therefore are independent of $c_0$. We conclude that
in the general case, \(TT) is replaced by 
$$
\langle T_0(z)T_0(0)\rangle=\left[c_0+\left({1\over(1-2\pi\lambda k)^2}
-1\right)\right]{1\over2z^4}
\eqno(TTT)
$$
However, the other correlation functions $\langle T_0\no{J^2}\rangle$
and $\langle\no{J^2}\no{J^2}\rangle$ involve only $k$ and not $c_0$.
As a result, when the four contributions are summed, all the
$k$-dependent pieces cancel just as they do in the Gaussian theory,
with
the result that $c=c_0$. 

\head{3. Application to oriented SAWs.}

In this section, we apply these results to the complex $O(n)$ model.
On a lattice, this may be described by a partition function of the form
$$
Z={\rm Tr\,}\prod_{r,r'}\left(1+xs_a(r)s_a^*(r')+{\rm cc.}\right)
\eqno()
$$
where $s_a(r)$ is a complex $n$-component spin at the site $r$,
normalized so that ${\rm Tr\,}s_a(r)s_b^*(r')=\delta_{ab}\delta_{rr'}$,
and the product is over nearest neighbour pairs $(r,r')$.
When expanded in powers of $x$, $Z$ may be expressed as a sum over
configurations of oriented self-avoiding loops, with each link weighted
by a factor $x$ and each loops by a factor $n$. In the limit $n\to0$ we
have the problem of a single self-avoiding loop. This theory has an
obvious $U(1)$ symmetry under phase rotations $s_a\to e^{i\alpha}s_a$,
and the corresponding current is the lattice version of
$J_\mu=(i/2)(s_a^*\partial_\mu s_a-s_a\partial_\mu s_a^*)$. It
corresponds to a unit current along each loop in the direction of its
orientation. 

In addition to closed loops, one may also consider open oriented SAWs.
A source at which $q$ distinct SAWs originate is related in the $O(n)$
model to the insertion of an operator $\phi^{(q)}_{a_1\ldots a_q}\equiv
s_{a_1}\ldots s_{a_q}$, which has charge $q$ under this $U(1)$
symmetry. More precisely, the correlation function
$$
\langle \phi^{(q)}(0)s^*_{a_1}(r_1)\ldots s^*_{a_q}(r_q)\rangle
\eqno()
$$
gives the generating function for the number of `star' oriented
polymers
which originate at $r=0$ and have their ends at $r_1,\ldots,r_q$.
In the absence of any orientation-dependent interactions, the sum
over both orientations of each loop is equivalent to a problem of 
unoriented loops, and the complex $O(n)$ model is equivalent to a
real $O(2n)$ model. In that case, the critical limit of this model
is known to be described in the continuum limit by a conformal field
theory with $O(2n)$ symmetry, which may be derived by making an
appropriate modification of the Gaussian model, or Coulomb gas, with a
charge on the boundary. From this analysis, the scaling dimensions of the
operators $\phi^{(q)}$ are known to be\refto{SALEUR}
$$
x_q(0)= {gq^2\over 8}-{(g-1)^2\over2g}
\eqno()
$$
where $n=-\cos\pi g$ with $1\leq g\leq2$. 
This gives the exponent $\gamma_q=\nu(2-x_q(0)-qx_1(0)$ 
which appears in the asymptotic
behaviour of the number $c^{(q)}_N\sim N^{\gamma_q-1}\mu^N$
of such star polymers. Of course, $q=1$ corresponds to the usual
case of a linear polymer, with $\gamma_1=\gamma=\nu(2-\eta)$.

Now consider adding an interaction energy between different parts of 
loops which depends on their relative orientation. In general, this may
be written as
$$
\sum_{r,r'}F(r-r')J_\mu(r)J_\mu(r')
\eqno()
$$
where the function $F$ is assumed to be short-ranged. In describing
the continuum limit of the effects of this perturbation, we may use the
OPE 
$$
J_\mu(r)J_\mu(r')={\rm singular\ term\ } +\no{J\Jb}+\cdots
\eqno()
$$
The singular term leads to a trivial constant in the energy, and may be
subtracted off. The terms represented by dots are all irrelevant.
We are therefore left with precisely the type of perturbation described
in the preceding section,
with a bare coupling constant $\lambda_0=\sum_rF(r)$. It should be
stressed that this is not necessarily equal to the renormalized coupling
constant $\lambda$ introduced in \sec{2}, since it will be renormalised
in a non-universal way by the irrelevant couplings. We do expect
$\lambda_0$ and $\lambda$ to have the same sign, however.

The introduction of $\lambda_0$ will also, in general, give additional
mass renormalisation, and hence will shift the value of the bare mass at
which the renormalised mass vanishes. This corresponds to a shift
in the critical value $x_c$ of the fugacity $x$ of the lattice model, and
hence of the connective constant $\mu=x_c^{-1}$. We may avoid this
problem by considering the following special case. Consider SAW on the
square lattice. In this model, the walks are never allowed to traverse
the same edge more than once. However, 
it is possible to have configurations where opposite edges of an
elementary square are
occupied by different parts of the same walk (or, for $n\not=0$, by
parts of different walks.) Call this a close encounter.
When this happens, the opposite edges may be
traversed either in a parallel, or an anti-parallel, fashion (see
\fig{1}.) In the
ordinary SAW problem, each of these configurations is assigned unit
weight. Now consider giving each parallel close encounter a weight
$w=e^{-\lambda_0}\not=1$, meanwhile continuing to assign anti-parallel 
encounters unit weight. This will then correspond to an energy difference
of the kind we require, as well as an overall shift in the net repulsive
energy (an irrelevant perturbation.) It may be shown that
these two effects will cancel in
their contribution to mass renormalization, at least for $w<1$, and
hence the connective constant $\mu$ will be independent of $w$.
This will certainly be true for closed loops, which can contain no
parallel close approaches and are therefore insensitive to $w$. For
single open walks, let us define $c_{N,M}$ to be the number of such
walks with $N$ steps and $M$ such parallel close encounters.
Thus, we
are interested in the asymptotic behaviour of
$c_N(w)=\sum_Mc_{N,M}w^M$.
The following argument is due to A.~Owczarek\refto{AJG}
Clearly $c_{N,0}<c_N(w)<c_N(1)$ for $0<w<1$. Both of these bounds
increase as $\mu^N$ (apart from powers of $N$.) This is true of
$c_N(1)$ by definition. The lower bound is itself bounded below
by the number of walks which return to the vicinity of the origin
This is essentially the number of closed loops of length $N$,
which are known to have the same connective constant $\mu$. A similar
argument should also hold for star polymers, at least for $w<1$. The
attractive case $w>1$ is less clear. For sufficiently large $w$ we
expect that bound states can form between two or more parallel walks. 
In that case, the critical point will occur when the bound state
mass vanishes rather than that of a single particle, 
and this may occur at a smaller value of $x_c$.
Certainly, for sufficiently large $w$, much greater than $\mu$, the
dominant configurations of single walks will be those wrapped up as
tightly as possible to maximise the number of parallel close encounters. 
In that
case $c_N(w)$ will behave roughly as $w^N$, and the walks will be
compact.

From the above discussion, we see that, at least for $w<1$, and perhaps
for a small range of $w>1$, we may apply the results of \sec{2} to show
that the number of star polymers, weighted by $w$, behaves
asymptotically as
$$
c^{(q)}_N(w)\sim N^{\gamma_q(w)-1}\mu^N
\eqno(CW)
$$
where $\gamma_q(w)=\nu\big(2-x_q(\lambda)-qx_1(\lambda)\big)
$, and $x_q(\lambda)$ is given by formula \(XL).
Note that the energy operator $s_a^*s_a$ has $q=0$ and is therefore
insensitive to $\lambda$. Thus the fractal dimension of these walks is
the same as the ordinary case, with $\nu=\frac34$.

For the complex $O(n)$ model
the $U(1)$ anomaly number $k(n)$ has been calculated by Coulomb gas
methods to be\ref{AREA}
$$
k(n)={2\pi(g-1)\cot\pi g\over g}=
{2\pi n\arccos n\over\sqrt{1-n^2}(\pi+\arccos n)}
\eqno(KN)
$$
where $0\leq\arccos n\leq\pi$. (The normalization of $k$ differs from
that used in \ref{AREA} by a factor $\pi^2$.)
Note that for $n=0$, $k$ vanishes, so that \(XL) simplifies to
$$
x_q(\lambda)=\left(\frac9{48}+2\pi\lambda\right)q^2-\frac1{12}
\eqno()
$$
This then leads to \(GQ) of the Introduction.
For $\lambda>0$, corresponding to the repulsive case $w<1$,
the value of $x_q$ increases and therefore $\gamma_q$ decreases as
expected. It is interesting to note that, for the opposite case
$\lambda<0$, there
appears to be a value of $\lambda$ for which $x_1$ vanishes. It is
tempting to suppose that this corresponds to the collapse transition
where the single walk winds up on itself, but there is no real evidence
for this. Also, because of the renormalization effects, we do not know
whether this critical value of $\lambda$ occurs for a finite value of
the bare coupling $\lambda_0$.
In fact this change of sign of the scaling dimensions for sufficiently
large $\lambda$ may be shown to occur for all values of $n\leq1$.

An interesting result follows if we differentiate \(CW) with respect to
$w$ and set $w=1$. This gives an estimate for the average number of
of parallel close encounters of an oriented star polymer:
$$
\bar M_N={\sum_MMc_{N,M}\over\sum_Mc_{N,M}}\sim Aq^2\ln N
\eqno()
$$
where $A$ is non-universal, but independent of $q$. Since the mean
number of both types of encounter increases linearly with $N$,
this suggests that
these effects might be rather hard to see in enumerations with small
values of $N$.

\head{Conclusions.}

In this paper we have pointed out that the existence of
a non-trivial moduli space, or continuum of fixed points, is 
not restricted to conformal field theories with $c\geq1$, but already
occurs in non-unitary $c<1$ theories of interest in statistical
mechanics. All that is required is that the theory possess a conserved 
$U(1)$ current. Since all values of $c>0$ may be reached by composing 
$c<1$ theories, it follows that continuously varying exponents may 
occur for all positive values of the central charge.

We have pointed out a specific application to oriented SAWs with
orientation-dependent interactions. While our conclusions apply mainly
to the case where the forces are repulsive, these walks appear also to
undergo a new kind of collapse transition if the forces between parallel
segments of the walk are sufficiently attractive. This is not the usual
theta-point because it does not occur for closed loops. A similar kind
of collapse transition has recently been observed in an exactly solvable
lattice model of walks coupled to an Ising degree of freedom\refto{NPR}.
Note that the occurrence of what we have termed parallel close encounters
in a SAW implies that at least one end is `trapped' by the walk. Thus
the number of such trapped walks is thus given by $c_N(1)-c_N(0)$, and our
results indicate that a fraction $1-O(N^{-\gamma(0)+\gamma(1)})$
are trapped. Note that this exponent is not expected to be universal
since it depends on the precise definition of trapping.
While the occurrence of continuously varying exponents in the context
of oriented polymers is interesting, it should be noted that this effect
is peculiar to two dimensions. In $d>2$ dimensions, the ${\vec J}^{\,2}$
interaction has dimension $({\rm length})^{2-2d}$, and so is irrelevant.

The author acknowledges useful correspondence with A.~J.~Guttmann. 
This work was supported in part by a grant from the SERC.
 
\head{Figure captions.}

\item{1.} An oriented self-avoiding walk. $a$ indicates a parallel close
encounter; $b$ indicates an anti-parallel one. Note that only the latter
are allowed for close self-avoiding loops. In the model under
consideration, these two types of close encounter are assigned different
weights.

\item{2.} Some of the contributions to the $O(\lambda^2)$ correction to
the two-point function $\langle V_qV_{-q}\rangle$. In these diagrams,
the dashed line represents the unperturbed correlation function, solid
lines represent OPE contractions of the current with $V_{\pm q}$, and
the dotted line is a current-current contraction, proportional to $k$.
\references

\refis{FQS} D.~Friedan, Z.~Qiu and S.~Shenker,
\journal Phys. Rev. Lett., 52, 1575, 1984.

\refis{NIEN}  B.~Nienhuis, in {\sl Phase Transitions and Critical
Phenomena}, v. 11, C.~Domb and J.~L.~Lebowitz, eds. (Academic, 1986.)

\refis{MILL} J.~Miller, \journal J. Stat. Phys., 63, 89, 1991.

\refis{ZAM} A.~B.~Zamolodchikov, \journal Zh. Eksp. Teor. Fiz., 43, 565,
1986; [\journal JETP Lett., 43, 730, 1986.]

\refis{NPR} B.~Nienhuis, private communication.

\refis{AREA} J.~L.~Cardy, Oxford preprint OUTP-93-34S.

\refis{SALEUR} H.~Saleur, \journal J. Phys., A19, L807, 1986.

\refis{CS} S.~Chaudhuri and J.~A.~Schwartz, \journal Phys. Lett. B,
219, 291, 1989.

\refis{AJG} A.~J.~Guttmann, private communication.

\endreferences

\endit

#!/bin/csh -f
# Note: this uuencoded compressed tar file created by csh script  uufiles
# if you are on a unix machine this file will unpack itself:
# just strip off any mail header and call resulting file, e.g., cont.uu
# (uudecode will ignore these header lines and search for the begin line below)
# then say        csh cont.uu
# if you are not on a unix machine, you should explicitly execute the commands:
#    uudecode cont.uu;   uncompress cont.tar.Z;   tar -xvf cont.tar
#
uudecode $0
chmod 644 cont.tar.Z
zcat cont.tar.Z | tar -xvf -
rm $0 cont.tar.Z
exit
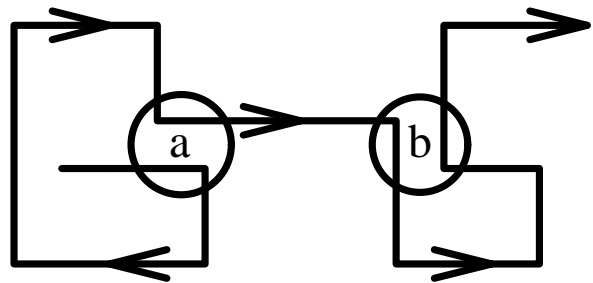
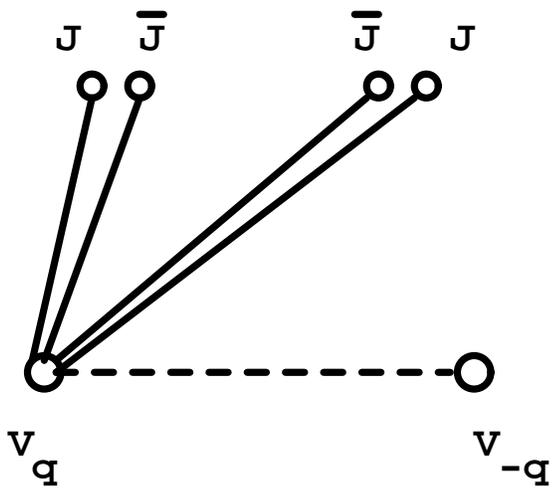
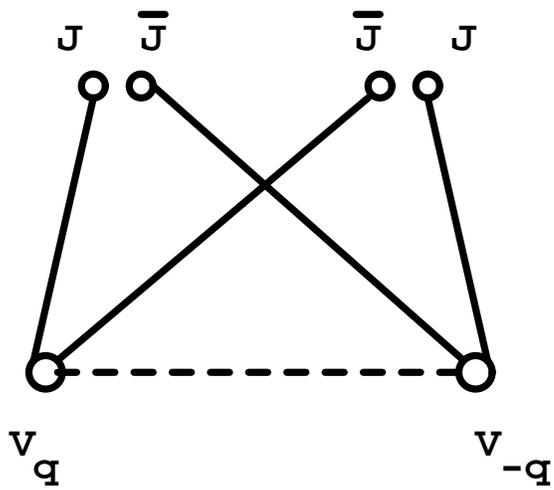
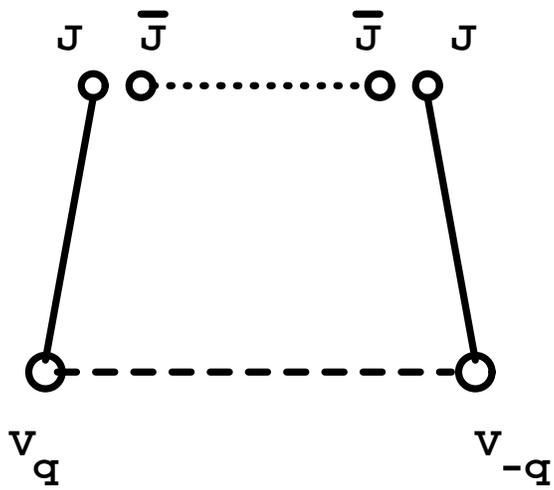
